\documentclass[]{aa} 
 
\usepackage{graphicx} 
\usepackage{amssymb} 
\usepackage{epsfig} 
\usepackage{natbib} 

\def\bb{{\sc bb}}

\def\nh{{$N_{\rm H}$}}  
 
\def\bb{{\sc bb}} 
\def\I{{\em INTEGRAL}} 
\def\S{{\em Swift}} 
\def\R{{\em RXTE}} 
\def\XTE{XTE~J1701-407}  
\def\be{\begin{equation}} 
\def\ee{\end{equation}}

\begin{document} 
 
\title{The new intermediate long bursting source \XTE\ } 
 
\author{M. Falanga,\inst{1,2} 
A. Cumming,\inst{3}
E. Bozzo,\inst{4,5} 
J. Chenevez\inst{6}  
} 
 
\offprints{M. Falanga} 
\titlerunning{Intermediate long bursts from XTE~J1701-407} 
\authorrunning{M. Falanga, et al.} 
 
\institute{CEA Saclay, DSM/IRFU/Service d'Astrophysique 
     (CNRS FRE 2591), 91191, Gif sur Yvette, France 
     \email{mfalanga@cea.fr} 
\and AIM - Unit\'e Mixte de Recherche CEA - CNRS - 
  Universit\'e Paris 7, Paris, France 
\and Physics Department, McGill University, 3600 rue University,
     Montreal QC, H3A 2T8, Canada
\and INAF - Osservatorio Astronomico di Roma, Via Frascati 33, 00044 
Rome, Italy 
\and Dipartimento di Fisica - Universit\'a di Roma Tor Vergata, via 
della Ricerca Scientifica 1, 00133 Rome, Italy 
\and National Space Institute, Technical University of Denmark, 
           Juliane Maries Vej 30, 2100 
           CopenhagenØ, Denmark 
}

\abstract 
{} 
{\XTE\ is a newly discovered X-ray transient source. In this work we
  investigate its flux variability and study the
  intermediate long and short bursts discovered by \S\ on July
  17, and 27, 2008, respectively.}   
{So far, only one intermediate long burst, with a duration of
  $\approx$18~minutes and ten days later a short burst, have been
  recorded from \XTE.  
We analyzed the public available data from \S\ and \R,\ and compared 
the observed properties of the intermediate long burst
with theoretical ignition condition and light curves to investigate the possible nuclear
burning processes.   
}   
{The intermediate long burst may have exhibited a photospheric radius expansion,
allowing us to derive the source distance at 6.2~kpc assuming the
empirically derived Eddington luminosity for pure helium.  
The intermediate long burst decay was best fit by using two exponential
functions with e-folding times of $\tau_1=40\pm3$~s and $\tau_2=221\pm9$~s. 
The bursts occurred at a persistent luminosity of $L_{\rm
  per}=8.3\times10^{36}$~erg~s$^{-1}$ ($\approx$2.2\% of the Eddington
luminosity).  
For the intermediate long burst the mass accretion rate per unit area
onto the NS was 
$\dot{m}\approx4\times10^{3}$~g~cm$^{-2}$~s$^{-1}$, and the total energy
released was $E_{\rm burst}\approx3.5\times10^{40}$~erg. This corresponds
to an ignition column depth of $y_{\rm ign}\approx1.8\times10^{9}$~g~cm$^{-2}$,
for a pure helium burning. We find that the energetics of this burst
can be modeled in different   
ways, as (i) pure helium ignition, as the result of either pure helium  
accretion or depletion of hydrogen by steady burning during  
accumulation, or (ii) as ignition of a thick layer of hydrogen-rich  
material in a source with low metallicity. However, comparison of the  
burst duration with model light curves suggests that hydrogen burning  
plays a role during the burst, and therefore this source is a low  
accretion rate burster with a low metallicity in the accreted  
material. 
} 
{}

 
\maketitle 
 
\section{Introduction} 
\label{sec:intro} 

Type I X-ray bursts are among the most evident signatures that testify 
the presence of a neutron star (NS) in low mass X-ray binaries. 
These bursts are thermonuclear explosions that occur on the surface of
accreting NSs and are triggered by unstable hydrogen and/or helium burning 
\citep[see, e.g.,][for reviews]{lewin93,sb06}. 
Type I X-ray bursts were predicted theoretically by \citet{hansen75},
and several thousand bursts have been observed to date 
\citep[see, e.g.,][]{cornelisse03,galloway06,chelovekov06}. 
From the duration of the bursts measured by their decay parameter
$\tau$ \citep[see e.g.,][]{galloway06} three main branches are distinguished:
normal bursts, intermediate long bursts, and superbursts 
\citep[see Fig. 7 in][and references therein]{Falanga08}. These bursts
can be described by different fuel types, accretion rates, and they,
therefore,  also displays different recurrence times
\citep[e.g.,][]{sb06,cm04,galloway06}. 
Thanks to the long Galactic plane scan carried out with {\it BeppoSAX}, \I, and 
\S\ a number of rare intermediate long bursts and superbursts have
been observed.  
In most cases the rise of the burst is of $\approx1$ s, whereas the decay is 
approximately exponential, with a duration of a few seconds for normal bursts, 
tens of minutes for intermediate long bursts, and up to several hours for
superbursts  
\citep[e.g.,][]{ek04,intZ04,molkov05,intZ05,chenevez06,chenevez07,Falanga08}.  
The recurrence time of type I X-ray busts ranges from few hours to years, depending 
on the nuclear reactions involved \citep[see, e.g.,][for
  reviews]{lewin93,sb06}.

In this paper, we report on an intermediate long burst from the
X-ray transient \XTE. This source was discovered by the {\it
  Rossi X-ray Timing Explorer} (\R)  
during a routine Galactic bulge scan on June 8, 2008 \citep{markwardt08a}. 
At the time of the discovery, the source spectrum was best fit
with an absorbed power-law model with photon index $\approx2.2$ and a
neutral absorption column of  $\approx3.4\times 10^{22}$ cm$^{-2}$,  
the measured flux was $\approx 1.4\times 10^{-10}$ erg cm$^{-2}$ s$^{-1}$
in the 2-10 keV band \citep{markwardt08a}. \S/XTE follow-up
observations on June 11, 2008 
found both spectral and source flux measurements consistent with
the \R/PCA  results \citep{degenaar08}.  
On July 17, 2008, the \S/BAT camera detected a short flare consistent with the
position of \XTE, and $\sim 97$ s later \S/XRT measured a decaying
X-ray flux \citep{barthelmy08}.  Based on this BAT data the X-ray
spectrum was consistent with those observed for thermonuclear X-ray
bursts, therefore, this flare was associated to a most likely type I
burst \citep{markwardt08b}. 
Moreover, based on the \S/XRT time resolved spectrum during the flux
decay, \citet{linares08} measured the cooling tail and confirmed the
flare to be an intermediate long type I X-ray
burst. This led to the classification of \XTE\ as a NS low-mass X-ray 
binary \citep{linares08,linares08b}.
Ten days later, i.e., on July 27, BAT observed in the 15--150 keV band a
short bursts lasting $\approx10$~s \citep{sakamoto08}. 
Kilohertz quasi-periodic oscillations were observed at $\approx1150$~Hz  
\citep[see][for details]{strohmayer08}. The most accurate 
position of the source was provided by \S\ at 
$\alpha_{\rm J2000}$=17$^{\rm h}$01$^{\rm m}$44$\fs$3 and 
$\delta_{\rm J2000}$=-40${\degr}$51$\arcmin$29$\farcs$9 with an
estimated accuracy of 1$\farcs$7 \citep{starling08}. 
No infrared counterpart candidate was found at this position \citep{kaplan08}.

\section{Data Analysis and results} 
\label{sec:analysis} 
 
\begin{figure} 
\centerline{\epsfig{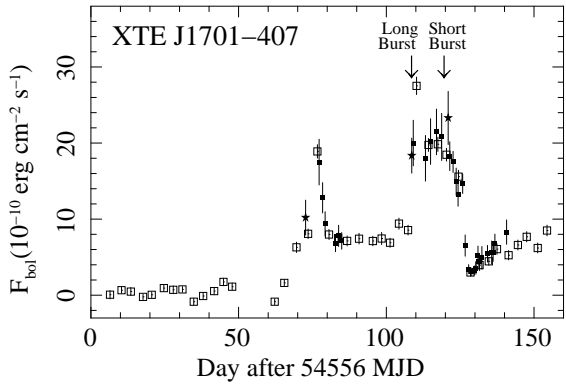}} 
\caption{Flux history of the X-ray transient \XTE. 
Open squares represent fluxes derived from 
{\it RXTE} bulge observations in the period from 
March 31, to Septembre 2, 2008. 
Fluxes derived from \R/PCA and \S/XRT target of opportunity and 
follow-up observations are 
marked with filled squares and stars, respectively. 
The arrow indicates the time of the type I X-ray burst.} 
\label{fig:fig1} 
\end{figure} 

For the present study we used  public available data from \S\ and
\R\ observatories. The dataset include \R\ target of 
opportunity observations, performed after the discovery of the source 
(observation ID. 93444), as well as \S\ follow-up observations 
(ID. 0031221001, 0031221002, and 00317205001). The \S\ data set
includes also the intermediate long burst discovered on July 17,
(ID. 00317205000) and the short burst from July 27, 2008 (ID. 00318166000). 
The total effective exposure time is 75.4 ks (30 pointings) and 1.1 ks
for \R/PCA and \S\, respectively. In Table~\ref{table:log} we report
the detailed observations log.
Data reduction of the \R\  Proportional Counter Array 
\citep[PCA; 2--60 keV,][]{jahoda96} and the High Energy X-ray Timing 
Experiment \citep[HEXTE; 15--250 keV,][]{rothschild98} was performed according  
to the default selection criteria for background subtraction, light curve, and 
spectrum extraction. We used only PCU2 data. 
We used \S/XRT data in both Windowed Timing (WT) and Photon
Counting (PC) mode. However, in the following we report only
on WT data, since we found that PC data were strongly affected by
pile-up and quasi-simultaneous observations carried out in WT mode
were available. We reduced all the XRT data by using the version of
the xrtpipeline (version~0.11.6)  included in
the Heasoft package 6.4, and the latest calibration files available.
Source events in WT mode
were extracted from rectangle regions with widths of 40 pixels and
heights of 20
pixels\footnote{See also the XRT analysis manual at
http://swift.gsfc.nasa.gov/docs/swift/analysis/xrt\_swguide\_v1\_2.pdf}.
Ancillary response files were generated with {\tt xrtmkarf} and
accounted for different extraction regions, vignetting, and
point-spread function. 
 For the \S/BAT data analysis we used the {\tt batgrbproduct} tool included
in the Heasoft package 6.4. Time resolved spectra and lightcurves (see
Sec. \ref{sec:lcburst})
were extracted by using the {\tt batbinevt}, {\tt batupdatephakw},
{\tt batphasyserr}, and {\tt batdrmgen} tools.

\subsection{Persistent emission}  
 \label{sec:spectra} 

Spectra obtained from \R/PCA and \S/XRT observations were fit
separately by using a photoelectrically-absorbed power-law
model. The best fit parameters for the absorption column, \nh, are
between $1.7\pm1$ and 
$4.4\pm1.2\times 10^{22}$ cm$^{-2}$ and a power-law index $\Gamma =
2.0\pm0.1-2.7\pm0.3$ the corresponding unabsorbed fluxes
are between $1.1\pm0.07\times10^{-10}$ and $8.3\pm0.6\times10^{-10}$
in the 2--20 keV energy band. In Table~\ref{table:log} we report
all the persistent spectral parameters. All uncertainties are at a 90\%
confidence level. PCA spectra were extracted in the 2--20~keV energy
band; the source was only detected at low significance in the HEXTE
(15--50~keV band). A fit to the joined broad-band PCA/HEXTE
(2--50~keV) spectrum did not lead to a significant improvement in the
determination of the model parameters; therefore, we report in the
following only on the PCA data. \S/XRT spectra were extracted in the
2--10~keV band. All the measured unabsorbed fluxes were extrapolated to the
0.3--100~keV band by generating dummy responses (XSPEC version
  11.3.2ag), and are shown in Fig.~\ref{fig:fig1}. We included
in this figure also fluxes derived from {\em RXTE} bulge
observations\footnote{http://lheawww.gsfc.nasa.gov/users/craigm/\\  
galscan/main.html} \citep{swankmark01}. In these cases, the conversion
between {\em RXTE} count rate and flux was obtained using the spectra results 
and the values reported by \citet{markwardt08a}.   
 
\begin{figure}[htb] 
\centerline{\epsfig{file=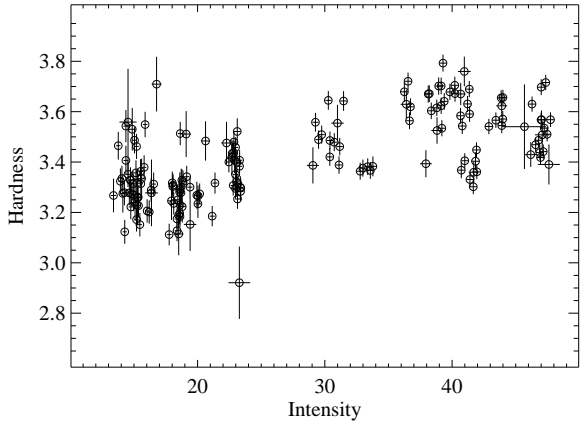,width=9.0cm}} 
\caption 
{HID of \XTE. The hardness is the ratio of the count rates in the
  5.7--18.9~keV to 2.1--5.7~keV and the intensity is in the 2.1--18.9
  keV count rate. Each point corresponds to 516~s of integration 
 time.}  
\label{fig:fig2} 
\end{figure}

The light curve and spectral analysis shows that \XTE\ 
underwent flaring activity lasting a few days, during which the
X-ray flux has varied by over a factor of 2--3. 
This variability can be ascribed to changes in the mass accretion 
rate or in the position of the source along its colour-colour 
diagram \citep[CCD, see e.g.,][]{hasinger89}. 
The latter possibility is investigated in Fig.~\ref{fig:fig2}. 
The CCD was realized by using background subtracted \R/PCA light 
curves with a 516~s time resolution. 
The soft colour is defined as the logarithm of the ratio between count
rates in the energy bands 2.1--3.7~keV and 3.7--5.7~keV, whereas for
the hard colour the energy bands 5.7--9.8~keV and 9.8--18.9~keV are used.  
No obvious transition between a hard to a soft state is observed. In order to reduce the errors on the 
colours, we generated a hard intensity diagram (HID) based on the net
count rates in the 2.1--5.7 keV and 5.7--18.9 keV with a 516 s time
bin. Howevere, the source behaviour is not much better traced in the
HID, because the statistical uncertainties are not significantly
reduced along the two axes. In Fig.~\ref{fig:fig2} we show the HID. 
Future observations are thus required in order to investigate further
the origin of flux variations in \XTE.\  
 
\begin{table*}[htb] 
\caption{Log of \R\ and \S\ observations and best fit 
  spectral parameters of the persistent emission. An absorbed power-law model is used to fit these spectra.} 
\begin{center} 
\begin{tabular}{ccccclllc} 
\noalign{\smallskip} 
\hline 
\hline 
\noalign{\smallskip} 
Instrument  &  Date  & Exp. & \nh\ & $\Gamma$ & Flux$_{\rm 2-10 keV}^{a}$ 
& Flux$_{\rm 2-20 keV}^{a}$ & Flux$_{\rm 0.3-100 keV}^{a}$ & $\chi^{2}_{red}$\\ 
&  MJD   &  ks  &  $10^{22}$ cm$^{-2}$ & & erg cm$^{-2}$ 
s$^{-1}$ & erg cm$^{-2}$ s$^{-1}$ & erg cm$^{-2}$ s$^{-1}$ &\\ 
\hline 
\noalign{\smallskip} 
\S$^{b}$ & 54628.66385 & $0.4$ & $4.0^{+0.7}_{-0.6}$ & $2.2^{+0.3}_{-0.3}$ 
& $2.8\pm0.3$e-10 & $3.7\pm0.5$e-10 & $10.2\pm2$e-10 & 0.8 \\  
\R\ & 54633.24637 & $2.0$ & $3.4^{+1.2}_{-2}$ & $2.4^{+0.1}_{-0.2}$ 
& $4.5\pm0.4$e-10 & $5.6\pm0.4$e-10 & $17.5\pm7$e-10 & 1.2 \\ 
\R\ & 54634.42452 & $2.5$ & $3.4^{+1.2}_{-1.2}$ & $2.4^{+0.1}_{-0.2}$ 
& $3.4\pm0.2$e-10 & $4.3\pm0.2$e-10 & $12.8\pm2$e-10 & 0.7 \\ 
\R\ & 54635.27545 & $2.0$ & $3.3^{+2.1}_{-2.2}$ & $2.2^{+0.1}_{-0.2}$ 
& $2.7\pm0.2$e-10 & $3.6\pm0.2$e-10 & $9.4\pm1.6$e-10 & 0.8 \\ 
\R\ & 54638.80878 & $3.5$ & $3.4^{+0.8}_{-0.9}$ & $2.2^{+0.1}_{-0.1}$ 
& $1.9\pm0.1$e-10 & $2.6\pm0.1$e-10 & $6.8\pm1.1$e-10 & 0.9 \\ 
\R\ & 54639.85563 & $3.0$ & $3.4^{+0.7}_{-0.7}$ & $2.3^{+0.1}_{-0.1}$ 
& $2.1\pm0.1$e-10 & $2.8\pm0.1$e-10 & $7.8\pm1.4$e-10 & 1.2 \\ 
\R\ & 54640.90230 & $3.5$ & $3.5^{+0.7}_{-0.7}$ & $2.2^{+0.1}_{-0.1}$ 
& $1.9\pm0.1$e-10 & $2.6\pm0.1$e-10 & $7.1\pm1.1$e-10 & 1.2 \\ 
\S$^{b,c}$ &  54664.62684 & $0.6$ & $3.1^{+0.4}_{-0.4}$ & $1.9^{+0.2}_{-0.2}$ 
& $4.9\pm0.3$e-10 & $7.2\pm0.6$e-10 & $18.3\pm3$e-10 & 1.1 \\ 
\R\ & 54665.10546 & $3.0$ & $3.7^{+1.2}_{-2}$ & $2.6^{+0.2}_{-0.1}$ 
& $4.9\pm0.6$e-10 & $5.9\pm0.7$e-10 & $20.0\pm4$e-10 & 1.6 \\ 
\R\ & 54669.24527 & $1.8$ & $2.9^{+1.2}_{-2}$ & $2.5^{+0.2}_{-0.2}$ 
& $4.8\pm0.8$e-10 & $5.8\pm0.9$e-10 & $18.0\pm3$e-10 & 1.6 \\ 
\R\ & 54670.99462 & $3.3$ & $3.9^{+1.2}_{-2}$ & $2.5^{+0.1}_{-0.2}$ 
& $5.5\pm0.7$e-10 & $6.5\pm0.6$e-10 & $23.2\pm4$e-10 & 1.5 \\ 
\R\ & 54672.97638 & $1.7$ & $4.1^{+1.2}_{-2}$ & $2.7^{+0.2}_{-0.2}$ 
& $4.9\pm0.5$e-10 & $5.8\pm0.6$e-10 & $21.5\pm5$e-10 & 1.3 \\ 
\R\ & 54674.73749 & $1.8$ & $4.2^{+1.2}_{-2}$ & $2.7^{+0.2}_{-0.3}$ 
& $4.6\pm0.7$e-10 & $5.5\pm0.6$e-10 & $20.8\pm5$e-10 & 1.6 \\ 
\S$^{b,d}$ &  54674.94054 & $0.1$ & $3.4^{+0.4}_{-0.4}$ & $2.0^{+0.4}_{-0.4}$ 
& $7.3\pm0.5$e-10 & $8.3\pm0.6$e-10 & $23.3\pm4$e-10 & 1.4 \\ 
\R\ & 54677.47471 & $3.5$ & $4.2^{+1.1}_{-1.2}$ & $2.6^{+0.3}_{-0.2}$ 
& $4.5\pm0.7$e-10 & $5.3\pm0.8$e-10 & $18.3\pm3$e-10 & 1.2 \\ 
\R\ & 54678.59119 & $3.3$ & $4.4^{+2.2}_{-1}$ & $2.3^{+0.3}_{-0.1}$ 
& $4.3\pm0.7$e-10 & $5.0\pm0.8$e-10 & $17.5\pm4$e-10 & 1.5 \\ 
\R\ & 54679.63915 & $3.2$ & $4.1^{+1.2}_{-2}$ & $2.7^{+0.2}_{-0.3}$ 
& $4.0\pm0.5$e-10 & $5.1\pm0.6$e-10 & $14.9\pm5$e-10 & 1.3 \\ 
\R\ & 54680.29712 & $2.1$ & $2.9^{+1.2}_{-1.4}$ & $2.5^{+0.2}_{-0.3}$ 
& $3.7\pm0.5$e-10 & $4.3\pm0.6$e-10 & $13.2\pm3$e-10 & 1.5 \\ 
\R\ & 54681.86101 & $3.5$ & $3.6^{+1.2}_{-2}$ & $2.3^{+0.2}_{-0.2}$ 
& $3.3\pm0.5$e-10 & $3.7\pm0.6$e-10 & $14.7\pm3$e-10 & 1.1 \\ 
\R\ & 54682.71212 & $7.0$ & $2.3^{+0.9}_{-0.9}$ & $2.5^{+0.1}_{-0.1}$ 
& $1.5\pm0.04$e-10 & $1.9\pm0.05$e-10 & $6.5\pm1.4$e-10 & 1.1 \\ 
\R\ & 54683.89045 & $3.5$ & $2.2^{+1.2}_{-1.2}$ & $2.2^{+0.1}_{-0.2}$ 
& $0.95\pm0.07$e-10 & $1.2\pm0.5$e-10 & $3.4\pm0.6$e-10 & 0.8 \\ 
\R\ & 54684.93786 & $1.8$ & $2.2^{+0.9}_{-1}$ & $2.2^{+0.2}_{-0.2}$ 
& $0.89\pm0.03$e-10 & $1.4\pm0.3$e-10 & $3.2\pm0.5$e-10 & 0.9 \\ 
\R\ & 54685.80434 & $1.6$ & $2.0^{+1.0}_{-0.6}$ & $2.2^{+0.1}_{-0.1}$ 
& $0.84\pm0.06$e-10 & $1.1\pm0.07$e-10 & $3.0\pm0.6$e-10 & 1.2 \\ 
\R\ & 54686.31211 & $2.5$ & $1.3^{+0.9}_{-0.4}$ & $2.3^{+0.1}_{-0.1}$ 
& $0.94\pm0.06$e-10 & $1.2\pm0.6$e-10 & $3.5\pm0.8$e-10 & 0.8 \\ 
\R\ & 54687.04156 & $0.7$ & $4.0^{+2}_{-2}$ & $2.5^{+0.2}_{-0.2}$ 
& $1.2\pm0.1$e-10 & $1.5\pm0.2$e-10 & $5.2\pm1.2$e-10 & 0.7 \\ 
\R\ & 54687.35989 & $2.7$ & $2.7^{+1.2}_{-1.2}$ & $2.3^{+0.1}_{-0.1}$ 
& $1.1\pm0.06$e-10 & $1.5\pm0.06$e-10 & $4.4\pm0.9$e-10 & 1.3 \\ 
\R\ & 54687.43638 & $2.2$ & $2.7^{+1.3}_{-1.3}$ & $2.3^{+0.1}_{-0.1}$ 
& $1.1\pm0.07$e-10 & $1.4\pm0.08$e-10 & $4.4\pm1.0$e-10 & 1.5 \\ 
\R\ & 54688.34471 & $2.4$ & $2.7^{+1.2}_{-1.2}$ & $2.4^{+0.1}_{-0.1}$ 
& $1.2\pm0.05$e-10 & $1.6\pm0.05$e-10 & $4.9\pm2$e-10 & 1.2 \\ 
\R\ & 54690.23786 & $2.2$ & $2.3^{+1.1}_{-1.0}$ & $2.3^{+0.1}_{-0.1}$ 
& $1.4\pm0.06$e-10 & $1.9\pm0.1$e-10 & $5.5\pm1$e-10 & 1.4 \\ 
\R\ & 54691.67712 & $1.8$ & $1.9^{+1.1}_{-1.0}$ & $2.3^{+0.1}_{-0.1}$ 
& $1.5\pm0.07$e-10 & $2.0\pm0.07$e-10 & $5.6\pm1.5$e-10 & 1.6 \\ 
\R\ & 54692.13508 & $1.6$ & $1.7^{+1}_{-1}$ & $2.2^{+0.1}_{-0.1}$ 
& $1.5\pm0.07$e-10 & $2.0\pm0.06$e-10 & $5.6\pm1.1$e-10 & 1.3 \\ 
\R\ & 54692.74785 & $1.2$ & $3.1^{+1.3}_{-1.3}$ & $2.3^{+0.1}_{-0.1}$ 
& $1.8\pm0.07$e-10 & $2.3\pm0.1$e-10 & $6.8\pm1.2$e-10 & 1.3 \\ 
\R\ & 54696.71397 & $0.5$ & $3.2^{+2}_{-1.5}$ & $2.3^{+0.1}_{-0.1}$ 
& $2.1\pm0.1$e-10 & $2.8\pm0.2$e-10 & $8.2\pm1.6$e-10 & 1.1 \\ 
\hline  
\end{tabular} 
${^a}$~Unabsorbed flux. $^{b}$~\S/XRT was operating in WT mode. $^{c,d}$
Intermediate long and short pre-burst persistent emission, respectively.  
\end{center} 
\label{table:log} 
\end{table*}

\subsection{Intermediate long burst light curves and spectra} 
\label{sec:lcburst} 
 
\begin{figure}[htb] 
\centerline{\epsfig{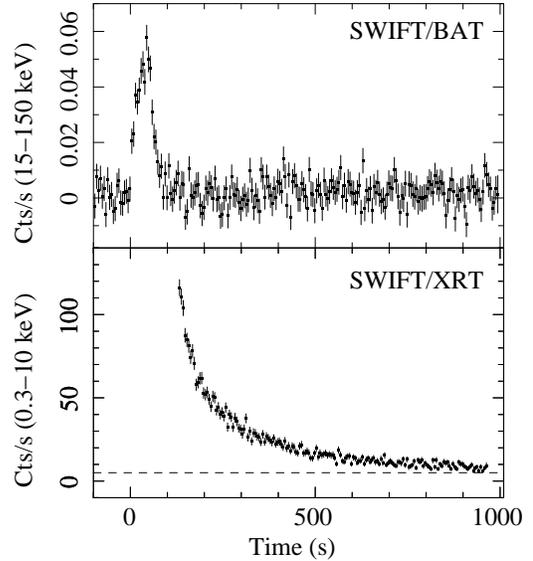}} 
\caption{The intermediate long type I X-ray bursts detected from \XTE\ on July 
17, 2008. The time T$_{0}$ expressed in UTC corresponds to $17^{\rm 
  h}31^{\rm  m}55^{\rm s}$. The \S/BAT (15--150 keV) and XRT (0.3--10 
keV) light curve are shown with a time bin of 5 s.  
The data gap of $\approx 133$ s in the XRT light curve is due to the 
time elapsed between the BAT trigger and the follow-up
observation. The dashed line indicates the background level measured
$\approx3000$~s after the burst end.  
}  
\label{fig:fig3} 
\end{figure}

In Fig. \ref{fig:fig3}, we show the \S/BAT 15--150 keV (upper panel) 
and XRT 0.3--10 keV (lower panel) light curve of the intermediate long burst. 
The burst start time, is the time at which
the BAT X-ray intensity of the source rose to 10\% of the peak above  
the persistent intensity level. The XRT light curve starts with a
delay of 133~s after the beginning of the burst as seen by BAT. The
BAT light curve shows a slow rise time\footnote{The rise time is
  defined as the time spent between the 
start of the burst and the point at which the 90\% of the peak burst
intensity is reached.} of $\approx45$~s.
The XRT decay time from the burst is best fit by using two exponential functions, 
with e-folding times of $\tau_1=44\pm8$~s and $\tau_2=271\pm21$~s,
respectively \citep[see also][]{linares08}.  
The total duration of the burst, i.e. the time spent to 
go from and back to the persistent state, was of $\approx86$~s and
$\approx18$~minutes in the BAT (15--150 keV) and the XRT (0.3--10
keV) light curves, respectively.     

\begin{figure}[htb] 
\centerline{\epsfig{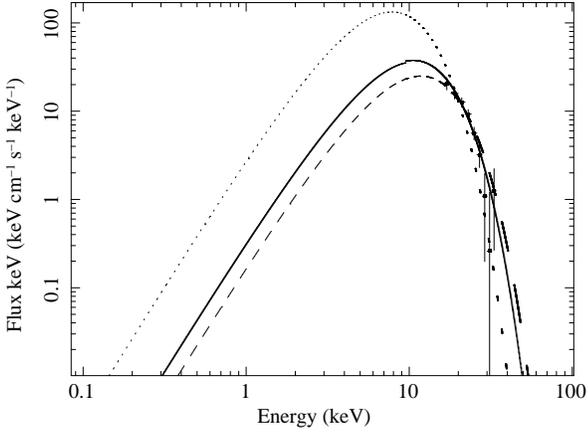}} 
\caption 
{\S/BAT (15--35 keV) data of the intermediat long peak black-body spectra
and best fit model, $kT_{\rm bb}\approx2.7$, (solid line). Upper limit,
$kT_{\rm bb}=3$ keV, acceptable fit (dashed-line) and lower,
$kT_{\rm bb}=2$, keV unacceptable fit (dotted-line).}   
\label{fig:figbb} 
\end{figure}

\begin{figure}[htb] 
\centerline{\epsfig{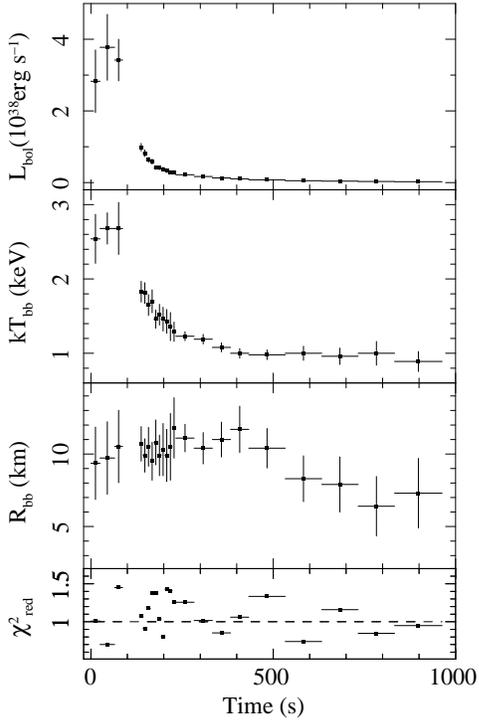}} 
\caption 
{Evolution of the spectral parameters, as inferred from \S/BAT (first
86~s) and \S/XRT (833~s) observations. The bolometric
luminosity is calculated by assuming a distance of 6.2 kpc, see
Sec. \ref{sec:3-1}.  
The bottom panel shows the $\chi^{2}_{\rm red}$
values for these fits.  
}  
\label{fig:fig4} 
\end{figure}

Type I X-ray bursts are produced by unstable burning of
accreted matter on the surface of the NS. The emission can
be well described by a black-body radiation with temperatures,
$kT_{\rm bb}$, in the energy range of a few keV. The energy dependent
decay time of these burts is attributed to the cooling of the NS
photosphere resulting in a gradual softning of the burst spectrum. For a review,
see, e.g. \citet{lewin93,sb06}.The time-resolved spectral analysis
of the burst was carried out by using  BAT and XRT data in the
15--35~keV and 0.3--10~keV bands, respectively. Burst spectra
were well fit by a simple photoelectrically-absorbed black-body,
\bb, model. For all these spectral fit the \nh value was frozen at
$3.1\times10^{22}$ cm$^{-2}$ derived from the pre-burst persistent
 emission (see Table~\ref{table:log}).
The inferred \bb\ temperature, $kT_{\rm bb}$,
apparent \bb\ radius at 6.2~kpc (see Sec. \ref{sec:3-1}), $R_{\rm bb}$, and
bolometric luminosity are shown in Fig. \ref{fig:fig4} and in Table
\ref{table:TB}.    
The burst fluence is calculated from bolometric 
fluxes, $F_{\rm bol}$; these correspond to the observed 2--10~keV \S/XTR fluxes 
extrapolated to the 0.1--100~keV energy range. 
The peak flux, $F_{\rm peak}$, is derived from the BAT 15--35~keV
light curve spectra and extrapolated to the 0.1--100~keV energy range. 
All the measured unabsorbed fluxes were extrapolated to the
0.1--100~keV band by generating dummy responses (XSPEC version
11.3.2ag). This is well justified for the XRT data since the
black-body temperature is well inside the spectra bandpass. However,
  during the main peak a black-body temperature reaches a maximun between
2--3 KeV \citep[see e.g.,][and references therein]{Falanga08}, and
therefore extrapolating the black-body spectra outside 
the BAT bandpass. For the BAT (15--35 keV) burst peak black-body spectral best
 fit we found $kT_{\rm bb}=2.68\pm0.21$ ($\chi^{2}=0.74$, 7 d.o.f.). We
fixed an upper and lower boundary black-body temperature of 2 and 3
keV, respectively, living the normalization as a free parameters. Note
the lower boundary is also consistent to be higher as the first XRT
black-body temperature of $kT_{\rm bb}\approx1.8$ (see Table~\ref{table:TB}). 
We found an unacceptable fit by using a 2 keV black-body burst peak temperature
($\chi^{2}_{\rm red}=5.1$, 8 d.o.f.), whereas a slightly better fit is
obtained by using a temperature of 3 keV ($\chi^{2}_{\rm red}=1.22$, 8
d.o.f.). In Fig.~\ref{fig:figbb} we show the
BAT data with the different black-body models from the different fits. In this
case the best fit values and the extrapolated peak flux are
acceptable values in the order of 20\% wich is inside our
BAT extrapolation error box. 

In Table~\ref{table:btab2} we report all the measured
bust parameters.  

\begin{table}[htb] 
\caption{Parameters of the Intermediate long burst.} 
\label{table:btab2} 
\begin{center} 
\renewcommand{\footnoterule}{} 
\begin{tabular}{ll} 
\hline \hline 
\noalign{\smallskip} 
$F_{\rm peak}^{a}$ (erg cm$^{-2}$ s$^{-1}$) &$ 8.2\pm2 \times10^{-8}$ \\ 
$f_{\rm b}^{b}$ (erg cm$^{-2}$)  & $7.6\pm0.3\times10^{-6}$ \\ 
$\tau \equiv f_{\rm b}/F_{\rm peak}$ (sec)  & $92\pm22$ \\ 
$\gamma  \equiv F^{c}_{\rm pers}/F_{\rm peak}$ & $2.2\pm0.5\times10^{-2}$\\ 
\hline 
\end{tabular} 
\end{center} 
\small $^{a}$Unabsorbed flux (0.1--100 keV). 
\small $^{b}$Fluence (0.1--100 keV). 
\small $^{c}$Unabsorbed persistent flux $F_{\rm
  pers}=(1.8\pm0.2)\times 10^{-9}$ erg cm$^{-2}$ s$^{-1}$ (0.1--100 keV).  
\end{table}

\begin{figure}[htb] 
\centerline{\epsfig{file=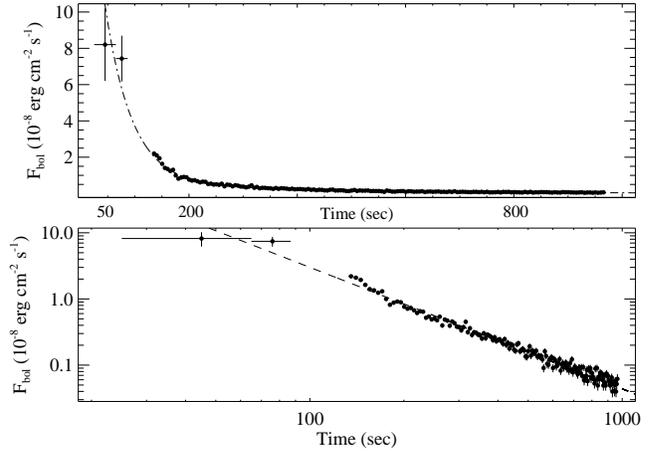,width=9.5cm}} 
\caption{Top panel: We show the BAT and XRT intermediate long burst
  light curve with the double exponential best fit curve. Bottom
  panel: The same data presented in a log-log scale with the power-law
  best fit model.   
}  
\label{fig:fig5} 
\end{figure}

From the time-resolved spectral results we converted the light curve
count rates to flux and found also in this case, that a double
exponential function are required to fit  
the intermediate long burst decay. The derived e-folding time are  
$\tau_1\approx40\pm3$~s and $\tau_2\approx221\pm9$~s, respectively with
a $\chi^{2}_{\rm red}= 1.1$ (for 163 d.o.f.). In this fit we included
also the first two BAT data points. Note, using only the XRT data
points we found the same results within the error bars. 
A single exponential function is found to be inadequate with
$\chi^{2}_{\rm red}=5.2$ (166 d.o.f.). In order to compare the decay
tail with the 
intermediate long X-ray burst from 2S 0918-549 and SLX 1737-282
\citep{intZ05,Falanga08}, and with the decay cooling model
from \citet{cm04} we fitted the data with a power-law, and found an
index of $\Gamma = -2.14\pm0.02$ with a $\chi^{2}_{\rm red}=1.7$ (167 d.o.f.). In
Fig. \ref{fig:fig5} we show the double exponential best fit (upper
panel) and in a log-log scale the power-law best fit (lower panel)
to the data. The double exponential fit is statistically
preferred over an power-law fit by the F-test probability of
$1.2\times10^{-14}$. Similar values of $\tau_1=44\pm7$~s and
$\tau_2=218\pm55$~s or power-law index $\Gamma = -1.97\pm0.09$ can be
obtained from the fit to the bolometric luminosity reported in 
Fig. \ref{fig:fig4}.   Also in this case the double exponential fit is
statistically  preferred over an power-law fit by the F-test probability of
$6.6\times10^{-3}$.

The decay time determined with the light curve in unit of count rates,
is restricted to the XRT 0.3--10 keV energy band. This decay time does
not take into account the different rise and cooling effect for different energy
bands during the outburst, where converting the rates to the bolometric flux or luminosity
using the time-resolved spectral results a more realistic decay time
can be obtained. Therefore, in the following we consider only the
$\tau_1\approx 40$ s and $\tau_2\approx 221$ s or the power-law
index $-2.14$. 
 In Table~\ref{table:btab2} we report the 
$\tau \equiv f_{\rm b}/F_{\rm peak}=92\pm22$~s which is only valid
if one exponential function describe the decay tail. In our case this value
is consistent within the error bars if we consider the double exponential decay as; $f_{\rm
  b}/F_{\rm peak} = \epsilon(\tau_1 -\tau_2) + \tau_2 = 71\pm12$,
where $\epsilon = F_{x}/F_{\rm peak}=0.83$ and $F_{x}$ is the peak
flux for $\tau_1$. 

\begin{table}[htb] 
\caption{Fit parameters of the time-averaged BAT and XRT burst spectra.} 
\begin{center} 
\begin{tabular}{ccccc} 
\noalign{\smallskip} 
\hline 
\hline 
\noalign{\smallskip} 
$\Delta$T  & $L_{\rm bol}$ & $kT_{\rm bb}$ &$R_{\rm bb}$ & $\chi^{2}$/d.o.f.\\ 
(s) & (10$^{38}$ erg s$^{-1}$) & (keV) & (km) & --\\
\hline 
\noalign{\smallskip} 
25$^{a}$  & 2.8$\pm$0.8      & 2.54$\pm$0.33 &  9.3$\pm$2.5 & 6.9/7\\ 
40$^{a}$  & 3.8$\pm$0.9      & 2.68$\pm$0.21 &  9.7$\pm$2.5 & 4.9/7\\
22$^{a}$  & 3.4$\pm$0.6      & 2.68$\pm$0.35 & 10.5$\pm$2.5 & 10.1/7\\
10  & 1.0$\pm$0.1      & 1.83$\pm$0.14 & 10.7$\pm$1.2 & 54.8/51\\
10  & 0.82$\pm$0.09    & 1.81$\pm$0.14 &  9.9$\pm$1.1 & 39.2/43\\
10  & 0.63$\pm$0.07    & 1.65$\pm$0.14 & 10.5$\pm$1.3 & 43.4/37\\
10  & 0.58$\pm$0.07    & 1.70$\pm$0.15 &  9.5$\pm$1.3 & 45.3/33\\
10  & 0.42$\pm$0.05    & 1.46$\pm$0.12 & 10.8$\pm$1.5 & 38.5/28\\
10  & 0.41$\pm$0.05    & 1.52$\pm$0.14 &  9.9$\pm$1.4 & 27.0/26\\
10  & 0.38$\pm$0.05    & 1.46$\pm$0.16 & 10.3$\pm$1.8 & 19.4/24\\
10  & 0.33$\pm$0.05    & 1.43$\pm$0.16 &  9.9$\pm$1.8 & 33.4/22\\
10  & 0.29$\pm$0.04    & 1.36$\pm$0.19 & 10.5$\pm$2.3 & 26.7/19\\
10  & 0.30$\pm$0.04    & 1.29$\pm$0.13 & 11.8$\pm$2.1 & 26.4/21\\
50  & 0.22$\pm$0.01    & 1.23$\pm$0.06 & 11.1$\pm$0.9 & 109.4/87\\
50  & 0.17$\pm$0.01    & 1.19$\pm$0.06 & 10.4$\pm$1.1 & 70.7/70\\
50  & 0.13$\pm$0.01    & 1.08$\pm$0.06 & 11.0$\pm$1.2 & 50.3/59\\
50  & 0.11$\pm$0.01    & 1.00$\pm$0.06 & 11.7$\pm$1.6 & 53.0/50\\
100 & 0.077$\pm$0.009  & 0.98$\pm$0.06 & 10.4$\pm$1.4 & 100.1/75\\
100 & 0.054$\pm$0.007  & 1.00$\pm$0.09 &  8.3$\pm$1.6 & 45.1/61\\
100 & 0.042$\pm$0.007  & 0.96$\pm$0.11 &  7.9$\pm$1.9 & 60.3/52\\
100 & 0.033$\pm$0.009  & 1.00$\pm$0.16 &  6.4$\pm$2.0 & 38.2/45\\
130 & 0.026$\pm$0.009  & 0.89$\pm$0.14 &  7.3$\pm$2.4 & 49.4/52\\
\hline  
\end{tabular}   
$^{a}$BAT spectrum.
\end{center} 
\label{table:TB} 
\end{table}

\subsection{Short burst light curves and spectra} 
\label{sec:lcburstshort} 
 
\begin{figure}[htb] 
\centerline{\epsfig{file=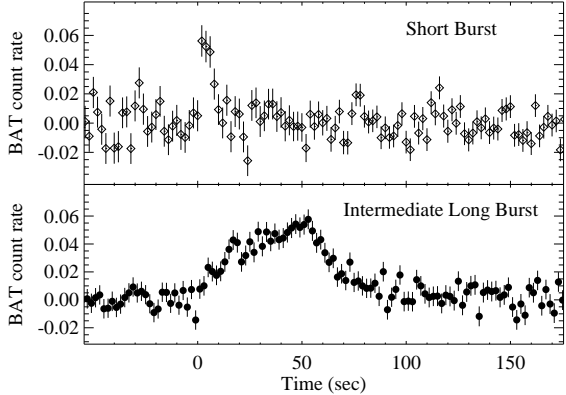,width=8.5cm}} 
\caption{Top panel: The short type I X-ray burst detected from \XTE\ on July 
27, 2008. For comparison of the rise time and duration we show also the
intermediate long burst (see also Fig. \ref{fig:fig3}). The \S/BAT
(15--150 keV) light curves are shown with a time bin of 2 s.   
}  
\label{fig:fig6} 
\end{figure} 

In Fig. \ref{fig:fig6} we show the \S/BAT
15--150 keV short burst light curve. The time T$_{0}$ expressed in UTC
corresponds to $22^{\rm  h}31^{\rm  m}19^{\rm s}$.
The rise time is $1\pm0.5$~s where the total duration is
$\approx8$~s. To determine the rise time and burst duration we rebined
the light curve to 1~s. For this short burst the XRT light curve starts with a
delay of 180~s after the beginning of the burst as seen by BAT. In the
XRT data (0.3--10 keV) the burst has not been detected, so the total
duration of this burst has to be considered $<180$~s. 
In  Fig. \ref{fig:fig6}  we show the short burst, and
for comparison we plot also the intermediate long burst seen by BAT.  
 
Given the short duration and statistic of the burst light curve we
were not able to study a time-resolved spectral analysis. 
Here we report the burst spectrum analysis integrated over 6 s in
order to measure as accurate as possible the peak flux.
Burst spectra were well fit by a simple black-body model. 
The inferred \bb\ temperature, $kT_{\rm bb}=3.9\pm0.5$~keV with an
unabsorbed bolometric peak fluxes, $F_{\rm
  bol}=3.8\pm1.1\times10^{-8}$ (0.1--100 keV). The burst occurred at a
flux persistent level of $2.3\pm0.4\times10^{-9}$ erg cm$^{-2}$
s$^{-1}$. This short burst occurred at a comparable persistent emission
level as the intermediate long burst.

\section{Discussion}  
\label{sec:discussion}

\subsection{Source distance, persistent flux, and accretion rate}
\label{sec:3-1} 

When a burst undergoes a photospheric radius expansion (PRE), the source
distance can be determined based on the assumption that the bolometric
peak luminosity is saturated at the Eddington limit, $L_{\rm Edd}$,
\citep[e.g.,][]{lewin93,kul03}. During the PRE episode, while the
bolometric luminosity remains constant at the Eddington 
value, the high energy flux may display a double peak profile and/or a
delay in the rise time  \citep[e.g.,][]{kul02,galloway06,Falanga07}.
The BAT light curve shows a slow rise time of $\approx45$~s, which is
typically observed at high energy in intermediate long helium bursts with PRE
\citep[e.g.,][]{kul02,molkov05,Falanga08}. 

Since \S/XRT missed the first 133 second of the burst from \XTE\ we
cannot directly determine whether this burst underwent a PRE. This issue cannot be
resolved with BAT time-resolved spectral analysis due to limited 
statistics (see Fig. \ref{fig:fig4}). However, the comparison with
other PRE bursts showing this slow rise time at high energy leads us
to take the observed profile in the BAT light curve as an evidence for
a PRE during the first $\approx 50$ s of the intermediate long burst, in
which case this should correspond to the timescale for the photosphere
to fall back to the neutron star surface. For the short burst no
conclusion can be drawn as to whether a PRE event occurred (see Fig.~\ref{fig:fig6}).  

Another possibility is that the burst has an intrinsically long rise
time. For example, the mixed H/He bursts observed from GS~1826-24
\citep{galloway04} do not show PRE, but have a rise lasting for 10
seconds set by hydrogen burning \citep{heger07}. The duration of the
bursts from GS~1826-24 is much less than the long burst from \XTE. It
is not clear whether a long duration mixed H/He burst could have a 50
second rise (see discussion of light curves in section 3.5).  
The fact that helium burning can be extremely rapid, whereas hydrogen
burning involves slow weak interactions means that the rise time is
longer when hydrogen is present, and PRE much less likely. For
example, \cite{fujimoto87} derived a critical helium fraction
necessary to achieve PRE. 

Assuming a bolometric peak luminosity equal to 
the Eddington value for a He type I X-ray burst
\citep[$L_{\rm Edd}\approx\,3.8\times 10^{38}$ ergs$^{-1}$, as
  empirically derived by][]{kul03},  
we obtain the source distance of $d=6.2^{+1.6}_{-0.9}$~kpc. 
For comparison, the theoretical value of this distance
\citep[e.g.,][]{lewin93} found by assuming a He atmosphere  
and canonical NS parameters (1.4 solar mass and radius of 10 km), is
5.5$^{+1.3}_{-0.8}$ kpc.  
Note that the source could be closer if the peak luminosity of the burst was lower than the pure helium Eddington limit. For example, we cannot rule out that the burst did not show PRE, in which case the peak luminosity could have been sub-Eddington. Alternatively, taking the peak luminosity to be the Eddington luminosity for solar composition ($X_0=0.7$) gives a distance of $\approx4$ kpc. In the following we consider $d\approx6.2$~kpc as a fiducial distance, and comment on how our conclusions would change if the source was closer.

The best fit to the 2--20~keV persistent emission spectrum of 
\XTE\ required an absorbed simple power-law model with $\Gamma \approx2.1$.  
Assuming a distance of 6.2~kpc, the estimated intermediate long burst
pre-burst persistent unabsorbed flux between 0.1--100~keV, $F_{\rm
  pers}\approx~1.9\times 10^{-9}$ erg cm$^{-2}$ s$^{-1}$, translates
into a bolometric luminosity of $L_{\rm pers}\approx~8.3\times
10^{36}$ erg s$^{-1}$, or $\approx2.2\% L_{\rm Edd}$. 

The local accretion rate per unit area is then given by $L_{\rm pers}=4\pi R^2\dot
m(GM/R)/(1+z)$, or 
\begin{eqnarray}
\dot m & = & 4.0\times 10^3\ {\rm g\ cm^{-2}\ s^{-1}}  \nonumber\\
& & \left({R\over 11.2\ {\rm
    km}}\right)^{-1}\left({M\over
  1.4\ M_\odot}\right)^{-1}\left({d\over 6.2\ {\rm
    kpc}}\right)^2\left({1+z\over 1.26}\right). 
\end{eqnarray}
A convenient unit for accretion rate is the Eddington accretion
rate. Here, we define the local Eddington accretion rate as $\dot
m_{\rm Edd}\equiv 1.8\times 10^5\ {\rm g\ cm^{-2}\ s^{-1}}$, the local
accretion rate onto a $M=1.4\ M_\odot$, and $R=11.2\ {\rm km}$ neutron star
that gives an accretion luminosity equal to the empirically derived
Eddington luminosity $3.8\times 10^{38}\ {\rm erg\ s^{-1}}$ from
\citet{kul03}. This gives $\dot m/\dot m_{\rm Edd}=2.2$\%.

\subsection{The energy, ignition depth, and recurrence time of the long burst} 
\label{sec:3-2}
 
The observed energy of the long burst allows us to estimate the
ignition depth. The measured fluence 
of the burst is $f_b=7.6\times 10^{-6}\ {\rm erg\ cm^{-2}}$, giving a
burst energy release $E_{\rm burst}=4\pi d^2f_b=3.5\times 10^{40}\ {\rm
  ergs}\ (d/6.2\ {\rm kpc})^2$. The ignition depth is given by $E_{\rm
  burst}=4\pi R^2y_{\rm ign}Q_{\rm nuc}/(1+z)$, or 
\begin{eqnarray}
y_{\rm ign} &=& 1.8\times 10^9\ {\rm g\ cm^{-2}} \left({d\over 6.2\ {\rm kpc}}\right)^2\nonumber\\
& &  \left(\frac{Q_{\rm nuc}}{1.6\ {\rm MeV/nucleon}}\right)^{-1}\left(\frac{R}{11.2\ {\rm
    km}}\right)^{-2}\left(\frac{1+z}{1.26}\right).  
\end{eqnarray}
The value of $Q_{\rm nuc}\approx 1.6$ MeV corresponds to the nuclear
energy release per nucleon for complete burning of helium to iron
group elements. Including hydrogen with mass-weighted mean mass
fraction $\langle X\rangle$ gives $Q_{\rm nuc}\approx 1.6+4\langle
X\rangle$ MeV/nucleon \citep{galloway04}, where we include losses due to
neutrino emission following \citet{fujimoto87}. For $\langle
X\rangle=0.7$, the solar composition value, $Q_{\rm nuc}=4.4$ MeV/nucleon, and
$y_{\rm ign}=6.5\times 10^8\ {\rm g\ cm^{-2}}$. 

At an accretion rate of $4\times 10^3\ {\rm g\ cm^{-2}\ s^{-1}}$, the
recurrence time corresponding to a column depth of $y_{\rm
  ign}=1.8\times 10^{9}\ {\rm g\ cm^{-2}}$ (pure helium composition) is $\Delta t=(y_{\rm
  ign}/\dot m)(1+z)=6.6\ {\rm days}$, or for $y_{\rm
  ign}=6.5\times 10^{8}\ {\rm g\ cm^{-2}}$ (solar composition) is
$\Delta t=2.4$ days. These derived recurrence times are independent of the assumed distance. The intermediate long burst, as the first
observed burst, occured 40 days after the detection of \XTE. 
The effective exposure time on the source from the
beginning of the outburst to the intermediate long burst was 0.34
days, and 0.14 days between both bursts. The elapsed time on the
source is thus too short compared to the theoretically derived
recurrence time to allow us to get an observational measurement of the
recurrence time.

\subsection{Theoretical comparison with ignition models}
\label{sec:3-3} 

\begin{table*}[htb]
\caption{Type I X-ray burst ignition conditions$^{a}$} 
\label{table:model} 
\begin{center} 
\begin{tabular}{llllllllllll}
\hline \hline 
\noalign{\smallskip}
Model & $\dot m^b$ & $Z_{\rm CNO}$ & $X_0^c$ & $Q_b$ & $y_{{\rm
    ign},9}$ & $T_{{\rm ign},8}$ & $\langle X\rangle$ & $X_b$ & $Q_{\rm
    nuc}$ & $E_{40}$ & $\Delta t$ \\ 
 & $(\%\,\dot m_{\rm Edd})$ &  & & & & & & & & & (days)\\
\hline 
\multicolumn{12}{c}{Pure helium accretion$^{d}$}\\
\hline
1 & 2.2 & 0.02 & 0 & 0.5 & 1.8 & 1.4 & 0 & 0 & 1.6 & 3.4 & 6.5\\
2 & 2.2 & 0.02 & 0 & 0.3 & 7.4 & 1.2 & 0 & 0 & 1.6 & 14 & 27\\
3 & 2.2 & 0.02 & 0 & 0.7 & 0.84 & 1.5 & 0 & 0 & 1.6 & 1.6 & 3.1\\
\hline
\multicolumn{12}{c}{Accretion hydrogen rich material}\\
\hline
4 & 2.2 & 0.02 & 0.7 & 0.5 & 0.13 & 2.0 & 0.38 & 0.07 & 3.1 & 0.49 & 0.48\\
5 & 0.69 & 0.02 & 0.7 & 0.5 & 1.6 & 1.4 & 0.01 & 0 & 1.64 & 3.2 & 19\\
6 & 2.2 & 0.001 & 0.7 & 0.1 & 0.67 & 1.7 & 0.62 & 0.54 & 4.1 & 3.3 & 2.5\\
7 & 2.2 & 0.001 & 0.7 & 0.5 & 0.53 & 1.8 & 0.63 & 0.43 & 4.1 & 2.7 & 2.0\\
\hline
\end{tabular}
\end{center} 
$^{a}$ Models 1, 5, and 6 provide a good match to the
  observed burst energy of $3.5\times 10^{40}\ {\rm ergs}$. In
  addition, models 1 and 6 have an accretion rate that matches the
  value inferred from the persistent luminosity. \\
$^{b}$We define $\dot m_{\rm Edd}=1.8\times 10^5\ {\rm
    g\ cm^{-2}\ s^{-1}}$, the local accretion rate onto a
  $1.4\ M_\odot$, $R=11.2\ {\rm km}$ neutron star that gives an
  accretion luminosity equal to the empirically derived Eddington
  luminosity $3.8\times 10^{38}\ {\rm erg\ s^{-1}}$ from
  \citet{kul03}.\\  
$^{c}$The hydrogen mass fractions are: in the accreted material $X_0$,
  at the base of the layer at ignition $X_b$, and the mass-weighted
  mean value in the layer at ignition $\langle X\rangle$.\\
$^{d}$Note that the ignition conditions for pure helium accretion do
  not depend on the choice of $Z_{\rm CNO}$. 
\end{table*} 

To try to understand the nuclear burning processes responsible for the long burst, we compare the observed properties with type I X-ray burst ignition
models. The ignition conditions are calculated as described in \citet{CB00}, but we take $R=11.2\ {\rm km}$, $M=1.4\ M_\odot$,
and the energy release in hot CNO burning\footnote{Here we adopt the
  value of $E_H$ from \citet{wallace81}, which includes
  neutrino losses. These were not included by \citet{CB00}.} 
$E_H=6.0\times 10^{18}\ {\rm erg\ g^{-1}}$. The results are
shown in Table \ref{table:model} for several different choices of accreted hydrogen
fraction $X_0$, accretion rate $\dot m$ and flux from the crust
$Q_b$. The ignition conditions are calculated by modeling the
temperature profile of the accumulating fuel layer, and adjusting the
thickness of the layer  until a condition for thermal runaway is met
at the base. These models do not include the effects of previous
bursts on the ignition conditions (e.g.~\citealt{w04}), and we have
not included gravitational sedimentation which is important at low
accretion rates \citep{Peng}. In the following, we consider three
posibilities: 1) accretion of pure
helium, 2) hydrogen-rich matter at solar, and 3) low metallicity.

\subsubsection{Pure helium accretion} 
\label{sec:3-3-1} 

For pure helium accretion, the accumulating
fuel layer is heated from below by a flux $\dot mQ_b$ from the
crust, where $Q_b$ is the energy per accreted nucleon flowing outwards from the crust.
We set the accretion rate at the observed value $\dot m/\dot
m_{\rm Edd}=2.2$\%, and adjust the value of $Q_b$ until we find
ignition at the inferred column  for pure helium burning $y_ {\rm ign}=1.8\times 10^9\ {\rm g\ cm^{-2}}$
(and therefore obtain the observed burst energy). A flux from below
$Q_b=0.5$ MeV/nucleon  matches the observed burst energy at $2.2$\%
Eddington. We also include models with $Q_b=0.3$ and $0.7$ MeV/nucleon in Table
\ref{table:model} to show the sensitivity of the ignition depth to the
amount of heat from below. Note that since the combination $\dot mQ_b$
sets the total flux heating the layer, an increase or decrease in
$Q_b$ keeping $\dot m$ fixed is equivalent to increasing or decreasing
$\dot m$ with $Q_b$ fixed. Therefore, if the inferred accretion rate is
smaller by some factor, the value of $Q_b$ needed to match the
observed energy will be larger by the same factor. Similarly, a source distance less than $6.2\ {\rm kpc}$ can be accommodated by increasing $Q_b$.

The outwards flux expected from the crust depends on accretion rate
and the core neutrino emissivity \citep{brown00}, ranging from $\dot mQ_b\approx
0.1$ MeV per nucleon at high accretion rates up to $\dot mQ_b\approx 1.5$ MeV
per nucleon at low accretion rates. The value depends on how much of
the total $\approx 1.5$ MeV per nucleon heat released in the crust by
pycnonuclear and electron capture reactions \citep{haensel90,haensel08} 
is conducted into the core and released as neutrinos compared to
being conducted outwards. The number of $0.5$ MeV/nucleon that we infer here
is quite reasonable for an accretion rate of 2.2\% Eddington; for
example, the models calculated by \citet{cum06} for
persistently accreting sources have $Q_b=0.3$ to $0.9$ MeV/nucleon at this
accretion rate. \citet{galloway06b} modelled the X-ray bursts
observed from SAX~J1808.4-3658 and found $Q_b\approx 0.3$ MeV/nucleon for $\dot
m\approx 3$\% $\dot m_{\rm Edd}$. 

\subsubsection{Accretion of hydrogen rich matter with solar metallicity}
\label{sec:3-3-2} 

Next, we consider that the source is accreting hydrogen
rich matter with the solar hydrogen fraction $X_0=0.7$ and a
metallicity similar to solar, with mass fraction of CNO elements
$Z_{\rm CNO}=0.02$. In that case, the hydrogen burns at a fixed rate
during accumulation of the fuel layer, by the beta-limited hot CNO
cycle of \cite{hoylefowler65}. The hydrogen depletes at a column depth  
\begin{eqnarray}
y_{\rm d}  & = &1.5\times 10^8\ {\rm g\ cm^{-2}}  \nonumber\\
& & \left({\dot m\over
  0.022\dot m_{\rm Edd}}\right)\left({Z_{\rm CNO}\over
  0.02}\right)^{-1}\left({X_0\over 0.7}\right) 
\end{eqnarray}
smaller than the inferred ignition depth, so that a thick layer of
pure helium accumulates and ignites beneath a steady hydrogen burning
shell. Therefore, even though solar composition material is accreted,
the mean hydrogen fraction at ignition in the fuel layer is very
small, giving $Q_{\rm nuc}$ close to the value $1.6$ MeV/nucleon for pure
helium. 

For a CNO mass fraction $Z_{\rm CNO}=0.02$, the observed burst energy
is obtained for an accretion rate three times lower than inferred from
the observed X-ray luminosity $\dot m/\dot m_{\rm Edd}=0.69$\% (model 5
in Table \ref{table:model}). At an accretion rate of 2.2\% Eddington,
the burst energy is a factor of seven too small (model 4). Reducing the distance to the source does not help because it would change both the inferred $\dot m$ and burst energy by the same factor.

\subsubsection{Accretion of hydrogen rich matter with low metallicity}
\label{sec:3-3-3} 

In the third scenario, we consider that the material is hydrogen
rich, but with a low 
metallicity. Model 6 has a burst energy close to the observed energy
at the inferred accretion rate of 2.2\% Eddington, with a CNO mass
fraction $Z=10^{-3}$, roughly 10\% of the solar metallicity. In this
case, the amount of hot CNO burning is reduced substantially, so that
hydrogen permeates the entire fuel layer at ignition. Hot CNO burning
still operates at a low level and causes some preheating of the fuel
layer. The hydrogen increases the amount of nuclear release during the
burst, giving $Q_{\rm nuc}\approx 4$ MeV/nucleon, and an ignition depth three
times smaller than for the pure helium case. The low metallicity ignition
is less sensitive to $Q_b$ than the pure helium ignition. Model 7,
which has the same conditions as model 6, but with $Q_b=0.5$ rather
than $Q_b=0.1$, has a burst energy within almost 20\% of the observed
value. Similar to the pure helium ignitions, a closer distance can be accommodated by varying $Q_b$.

One caveat regarding the low metallicity ignition models is that there
can be substantial heating of the accumulating fuel layer because of
nuclear reactions associated with the ashes of previous
bursts. \citet{w04} found that the burst behavior at
accretion rates $\approx 0.1$ Eddington was insensitive to the
metallicity of the accreted material due to this effect. However, they
found that at lower accretion rates there was good agreement with the
ignition models presented here \citep[see][Table 9]{w04}. 

\subsubsection{Summary of ignition models}
\label{sec:3-3-4} 

We present three ignition models in Table \ref{table:model} 
that fit the observations. Models 1 and 6 have the correct burst energy and
accretion rate. They correspond to accretion of pure helium (model 1), for which
the layer is heated by the outwards flux from the neutron star crust,
and for accretion of hydrogen rich material with low metallicity (model 6), for
which a low level of hydrogen burning preheats the layer during
accumulation, but the hydrogen fraction at ignition is substantial and
makes a significant contribution to the burst energetics. Third, model
5 has the correct burst energy, but at an accretion rate three times
lower than observed. Given the uncertainties in translating the
observed X-ray luminosity to accretion rate, it seems worthwhile
considering this model further. In this model, the accreted
composition is hydrogen-rich with a solar metallicity. This leads to
depletion of the hydrogen by the hot CNO cycle and the build-up of a
thick layer of pure helium beneath the hydrogen shell. 

There is a fourth possibility, which is that the source is accreting
hydrogen-rich material, but the hydrogen burns unstably in a series of
short flashes. The helium produced in the short flashes builds up and
make a pure helium layer that ignites to give the long burst. This is
similar to model 5, but with unstable rather than stable hydrogen
burning. 

One way to distinguish the various possible explanations for the burst
energetics would be a recurrence time measurement: as shown in Table
\ref{table:model}, the different scenarios predict differences in recurrence time. In
fact, this is equivalent to a measurement of the $\alpha$ parameter\footnote{The quantity $\alpha$ is defined as the
  ratio of the total energy emitted in the persistent flux to that
  emitted in a burst,
  $\alpha=F_{\rm pers}\Delta t / f_{\rm b}$, where $\Delta t$ is the
  time interval between two bursts.}
for the bursts, which would indicate the fuel type
(e.g.~$\alpha\approx 40$ for solar hydrogen abundance as found for
example in GS~1826-24, \citealt{galloway04}). 

While this paper was in preparation, a similar analysis of the
intermediate burst from \XTE\ was carried out and reported in the
preprint by \citet{linares08b}. The bolometric peak flux for the long
burst and therefore the limits on the source distance are consistent
with the values we obtain here. However, the persistent luminosity and
burst energy given in \citet{linares08b} are a factor of two smaller
than our values, being restricted to the energy range $1$--$50$ keV,
whereas we have included a bolometric correction. These values are
used in the interpretation of the long burst, resulting in an
accretion rate of $\dot m=1.9\times 10^3\ {\rm
  g\ cm^{-2}\ s^{-1}}=1.1$\% $\dot m_{\rm Edd}$ and a burst energy of
$E_b=1.6\times 10^{40}\ {\rm ergs}$, both a factor of two 
smaller than the values we find in this paper. Repeating the ignition
calculations presented earlier for this lower accretion rate, we find
that accretion of solar 
composition material ($X_0=0.7$) gives a burst
energy of $E_b=1.5\times 10^{40}\ {\rm ergs}$ for $Q_b=0.1$~MeV/nucleon, and
$E_b=1.0\times 10^{40}\ {\rm ergs}$ for $Q_b=0.5$~MeV/nucleon. 
Therefore, accretion of solar composition  
naturally explains the burst energetics for these values of $E_b$ and
$\dot m$. For pure helium accretion at this rate, we find that
$Q_b=1.5$ MeV/nucleon is required to achieve a burst energy of $1.5\times
10^{40}\ {\rm ergs}$ (lower values of $Q_b$ result in a deeper
ignition and more energetic burst). Whereas this is a larger value
than expected at this accretion rate, it 
is within the range of the total energy released in the crust
\citep{haensel08}. In addition, since the flux heating the layer is
$\propto\dot mQ_b$, the requirements on $Q_b$ can be relaxed if the
true accretion rate is larger than assumed here. Therefore, we find
that the energetics argument given in \citet{linares08b} against
explaining this burst as pure helium is overstated.

\subsection{Constraints from the light curve}
\label{sec:3-4} 

The shape and duration of the light curve offer another way to
determine the composition of the fuel that burns during 
the burst. While we do not have models available with exactly the same burst energy, the
low accretion rate models from \citet{w04} are within a
factor of two to four in energy and accretion rate, and so we compare these models with the
observed light curve. These models have $\dot M=3.5\times
10^{-10}\ M_\odot\ {\rm yr^{-1}}$, corresponding to a local accretion
rate $\dot m=1$\% $\dot m_{\rm Edd}$.   

\begin{figure}[htb] 
\centerline{\epsfig{file=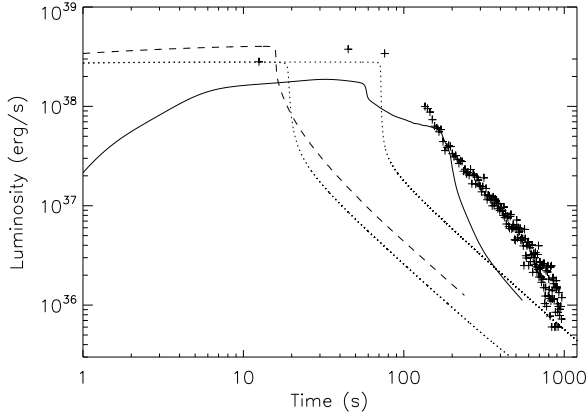,width=8.5cm}} 
\caption{Model light curves compared with the observed light curve. The solid 
  curve shows burst 2 from model zm of \citet{w04}. This burst has an
  ignition column 
  and energy a factor of 2 smaller than the observed burst. The dashed
  curve shows burst 3 from model Zm of \citet{w04}. This burst has an
  ignition column 
  a factor of three to four times smaller than the observed burst. The
  observed bolometric flux has been converted to luminosity using a
  distance of 6.2 kpc. Redshift corrections have been applied to the
  theoretical light curves, with $1+z=1.26$. The dotted curves show two
  cooling models calculated following \citet{cm04}, for
  (left to right) $y_{\rm ign}=6\times 10^8\ {\rm g\ cm^{-2}}$
  corresponding to Woosley et al.~(2004) model Zm (dashed curve), and
  $y_{\rm ign}=2\times 10^9\ {\rm g\ cm^{-2}}$ as needed to explain
  the observed burst energy (models 1 and 5 in Table \ref{table:model}). 
\label{fig:fig7}}
\end{figure}

First, we compare the observed light curve with burst 2 from model zm
of \citet{w04} (solid curve in Fig.~\ref{fig:fig7}). This model
is for accretion of hydrogen rich ($X_0\approx 0.7$) matter with a low
metallicity, 
$Z=10^{-3}$, and so is similar to model 6 in Table \ref{table:model}. 
The burst has a
total energy release of $2.0\times 10^{40}\ {\rm ergs}$, just less
than a factor of two smaller than the observed burst. The recurrence
time is $3.0$ days, and ignition column $y_{\rm ign}=3.6\times 10^8\ {\rm
  g\ cm^{-2}}$. The ignition column is just less than a factor of two
smaller than the observed burst. The peak luminosity of this
burst is close to the Eddington luminosity for solar composition,
suggesting that the distance may be closer than the $6.2\ {\rm kpc}$
assumed here. A closer distance would bring these light curves into
better agreement. The model light curve has a steep decline at late
times, somewhat steeper than the observed decline. An extra factor of
two in ignition column would give the correct burst energy and
lengthen the model burst light curve, bringing it into better agreement
with the observed burst. Another point to note is that this burst has
a slow rise time, lasting for several seconds, as expected for a low
helium mass fraction \citep{fujimoto87}. 

The dashed curve in Fig. \ref{fig:fig7} shows burst 3 from
model Zm of \citet{w04}. This model has hydrogen-rich matter with solar metallicity,
$Z=0.02$, and so is similar to model 5 of Table \ref{table:model}. 
The hydrogen burns
away in a thin shell, leaving a pure helium layer that ignites. This
burst has a total energy release of $8.0\times 10^{39}\ {\rm ergs}$, a
factor of 4 smaller than observed. The recurrence time is $4.5$ days,
and ignition column depth $5.6\times 10^8\ {\rm g\ cm^{-2}}$. The
burst reaches close to the pure helium Eddington luminosity. The
duration of the burst is shorter than the observed burst by a factor
of $\approx 5$--$10$. The slope of the decay is shallower than the
observed slope. The rise time of this burst is very fast, a fraction
of a second, in contrast to the much slower rise of the hydrogen-rich
burst (model zm; solid line in Fig.~\ref{fig:fig7}). 

We also include some cooling models calculated
following \citet{cm04}. For a given ignition column, an
energy release per gram of 1.6 MeV per nucleon is deposited in the
layer, as would be appropriate if the helium burned to iron group
elements at each depth at the start of the burst. We have also
computed models for a lighter ash and correspondingly smaller energy
deposition (\citealt{w04} find that the burning does not go all the
way to iron group in their model zM), but the differences are small,
and this does not change our conclusions. The cooling of the layer is
then followed, with the flux from the surface limited to the Eddington
luminosity for pure helium\footnote{As noted by \citet{w04}, the shape
  of the light curve as the luminosity begins to drop below the
  Eddington luminosity is likely not accurately reproduced by these
  models, which do not follow the outer layers in detail. We will
  improve our treatment of this in future work. However, we expect
  that the late time cooling is not sensitive to this.}. We show two
examples in Fig. \ref{fig:fig7}. The first has $y_{\rm 
  ign}=6\times 10^8\ {\rm g\ cm^{-2}}$ to match the \citet{w04} model
Zm burst. The second has $y_{\rm ign}=2\times
10^9\ {\rm g\ cm^{-2}}$ as needed to get the observed burst energy
(models 1 and 5 in Table \ref{table:model}). The shape of the cooling
models agrees well with the model Zm light curve, and agrees within a
factor of two in the cooling timescale. Even allowing for this factor
of two in the $y_{\rm ign}=2\times 10^9\ {\rm g\ cm^{-2}}$ model, the
cooling is faster than the observed light curve.  

In summary, although the models from \citet{w04} are not at  
exactly the same ignition conditions as implied by the observations of  
\XTE, our comparison suggests that pure helium ignition at  
the inferred ignition column depth (with or without a small overlying  
hydrogen burning shell) have cooling times that are shorter than the  
observed light curve. On the other hand, a hydrogen-rich composition  
throughout the layer, as expected for low metallicity, produces a  
longer lasting light curve more consistent with the observed cooling  
time. This conclusion does not depend on the assumed distance to the source. 
Both helium and hydrogen-rich burst models reach the Eddington  
luminosity (either the pure helium or solar composition Eddington  
luminosity respectively), and therefore could explain the PRE  
suggested by the similarity between the observed BAT light curve and  
other intermediate long bursts (see section 3.1). If the slow rise  
time is intrinsic to the burst and not due to PRE, the hydrogen rich  
model is preferred as the presence of hydrogen leads to a much slower  
rise time than for pure helium (Fig. \ref{fig:fig7}). The double
exponential nature   
of the decay may also argue for hydrogen burning during the burst. The  
burst profiles from GS~1826-24 \citep{galloway04} are well-fit by  
a double exponential decay. Further modeling is required to study the  
expected burst profiles of hydrogen-rich bursts produced as a result  
of low metallicity accretion.

\subsection{Origin of the short burst}
\label{sec:3-5} 

We have focused on the long duration burst, which has a well-measured
fluence and therefore energy. For the short burst, we can only place
an upper limit on its fluence. At low accretion rates, unstable
ignition of hydrogen can give rise to short duration
bursts \citep[e.g.,][]{sb06,chenevez07}. \citet{linares08b} suggest
that this is the origin of the 
short burst from \XTE, and that either (i) hydrogen-ignited short
bursts are producing the helium fuel for the long burst, or that (ii)
the source is accreting close to the boundary between unstable and
stable hydrogen burning, and stable hydrogen burning produces the
helium for the long burst. The comparison to ignition models and model
light curves that we have made earlier suggest that hydrogen survives
to the ignition depth, implying a low metallicity in the accreted
layer. Unfortunately, this would presumably make a thermal instability
driven by CNO burning less likely. 

For the low metallicity model or for pure helium accretion, another
explanation is that the local accretion rate was higher at the time of
the short burst. The ignition depth is very sensitive to the base flux
or equivalently to the product $\dot m Q_b$ (see for example Fig.~8 of
\citealt{intZ05}). The persistent flux at the time of the short burst
from \XTE\ was slightly larger than at the time of the long burst, even at 
30\%, not enough to cause a significant reduction in the ignition column 
depth. We note that short bursts were observed from
the intermediate long X-ray burster 2S~0918-549, a suspected
ultracompact binary and therefore accreting hydrogen-deficient
matter. Therefore, it is not clear to us that the observation of the
short burst rules out pure helium accretion in \XTE, as argued by
\citet{linares08b}.

\section{Conclusions}
\label{sec:conclusions} 

We have compared the observed properties of the long duration burst
from \XTE\ with models of type I X-ray burst ignition
conditions and light curves. We showed that the observed burst energy
could be understood as (i) pure helium ignition, either as a result of
pure helium accretion or of depletion of hydrogen by steady burning
during accumulation, or (ii) as ignition of a thick layer of
hydrogen-rich material with low metallicity. Comparing
with model light curves, we find that the pure helium ignitions cool
faster than observed. On the other hand, a hydrogen rich layer gives a
longer duration light curve with a steep decline in the tail of the
burst, better matching the observed light curve. Therefore we suggest that the  
intermediate long burst from \XTE\ was powered by unstable burning of  
a thick layer of hydrogen rich matter with low metallicity. 
Long X-ray bursts caused by pure helium ignitions beneath a hydrogen
shell have been identified, as for example by 
\citet{galloway06b} who argued that this was happening in
SAX~J1808.4-3658. But to our knowledge \XTE\ would be the first
example of a source that shows long bursts driven by a thick layer of
hydrogen-rich material. The bursts from GS~1826-24 are believed to be
powered by rp-process hydrogen burning, giving long $\approx 100$
second tails, but the ignition depth in those bursts is an order of
magnitude smaller than inferred for the long burst from \XTE,
so that hydrogen can survive until helium ignition, even for solar
metallicity. At the low accretion rate in \XTE, this is not
the case: low metallicity is required to reduce the rate of hot CNO
burning and allow hydrogen to survive until helium ignites.  
This implies either that this source is a burster accreting  
low metallicity H-rich material at a low rate, or another possibility  
is that heavy elements are able to sediment out from the accumulating  
layer at this low accretion rate \citep{Peng}, reducing its  
effective metallicity. If so, future studies of this source could be  
used to test the physics of sedimentation at low accretion rates.

\acknowledgements 
We thank Alexander Heger for providing the burst light curves from \citet{w04} shown in Fig. 7.
MF acknowledges the French Space Agency (CNES) for financial support. 
JC acknowledges financial support from ESA-PRODEX, Nr. 90057, and 
EB acknowledges ASI and MIUR. AC acknowledges support from the
National Sciences and Engineering Research Council of Canada (NSERC),
Le Fonds Qu\'eb\'ecois de la Recherche sur la Nature et les
Technologies (FQRNT), and the Canadian Institute for Advanced Research
(CIFAR). AC is an Alfred P. Sloan Research Fellow.

\end{document}